\documentclass[preprint,5p,times,compress]{elsarticle}

\usepackage{siunitx}
\usepackage{amsmath, amsthm, amssymb, graphicx, subfigure, braket, mhchem}
\usepackage[bookmarks=true, colorlinks, citecolor=blue, linkcolor=black]{hyperref}
\usepackage{color}

\DeclareMathAlphabet{\mathmybb}{U}{bbold}{m}{n}

\DeclareMathOperator{\Tr}{Tr}

\journal{Physics Letters A}

\begin{document}

\begin{frontmatter}

\title{Anomalous spin Josephson effect in spin superconductors}

\author[ND,CT]{W. Zeng\corref{cor1}}
\address[ND]{School of Physics and Electronic Engineering, Jiangsu University, Zhenjiang 212013, China}
\address[CT]{Jiangsu Engineering Research Center on Quantum Perception and Intelligent, Detection of Agricultural Information, Zhenjiang 212013, China}

\author[XD,XT]{R. Shen}
\address[XD]{National Laboratory of Solid State Microstructures and School of Physics, Nanjing University, Nanjing 210093, China}
\address[XT]{Collaborative Innovation Center of Advanced Microstructures, Nanjing University, Nanjing, 210093, China}

\cortext[cor1]{Corresponding authors. \\
\textit{E-mail address:} zeng@ujs.edu.cn}

\begin{abstract}
The spin superconductor state is the spin-polarized triplet exciton condensate, which can be viewed as a counterpart of the charge superconductor state. As an analogy of the charge Josephson effect, the spin Josephson effect can be generated in the spin superconductor/normal metal/spin superconductor junctions. Here we study the spin supercurrent in the Josephson junctions consisting of two spin superconductors with noncollinear spin polarizations. For the Josephson junctions with out-of-plane spin polarizations, the possible $\pi$-state spin supercurrent appears due to the Fermi momentum-splitting Andreev-like reflections at the normal metal/spin superconductor interfaces. For the Josephson junctions with in-plane spin polarizations, the anomalous spin supercurrent appears and is driven by the misorientation angle of the in-plane polarizations. The symmetry analysis shows that the appearance of the anomalous spin Josephson current is possible when the combined symmetry of the spin rotation and the time reversal is broken. 
\end{abstract}


\end{frontmatter}

\section{Introduction}\label{sec:1}
The exciton condensates are composed of the electron-hole pairs which are charge neutral while their spin may either be singlet or triplet \cite{PhysRevLett.124.166401,PhysRevB.104.195309}. The key issue to realize the stable exciton condensates is to prevent the electron-hole recombination \cite{tachiya2010theory,blom1997temperature}. Many strategies have been proposed to do so, such as the exciton condensates realized in the double-layer systems \cite{LOZOVIK1976391,PhysRevB.77.233405,PhysRevLett.101.246404,PhysRevB.84.155409}, where two monolayers are spatially separated by an insulator layer. The insulator is supposed to be high enough to prevent the electron-hole recombination. Recently, Sun \textit{et al.} \cite{PhysRevB.84.214501,PhysRevB.87.245427} suggested that the stable spin-polarized triplet exciton condensates could be realized in the graphene monolayer growing on a ferromagnetic material or under an external magnetic field, where the electron-like spin-up states are below the hole-like spin-down states \cite{PhysRevB.84.214501}, which prevents the electron-hole recombination and means the exciton condensates in the ferromagnetic graphene system are stable. These stable exciton condensates are termed as the spin superconductor states \cite{PhysRevB.84.214501,PhysRevB.87.245427}. Analogous to the Andreev reflection in charge superconductors \cite{pannetier2000andreev,PhysRevLett.81.3247,PhysRevLett.74.1657,PhysRevLett.103.237001,PhysRevLett.95.027002,PhysRevB.106.094503}, Lv~\textit{et al.} \cite{PhysRevB.95.104516} studied the Andreev-like reflection in the normal metal/spin superconductor (NS) junctions based on the ferromagnetic graphene, where an incident spin-up electron is reflected as a spin-down electron and the spin $2\times(\hbar/2)$ is injected into the spin superconductor. This spin-flipped Andreev-like reflection leads to the spin current when a spin voltage is applied across the junction.

% the Andreev-like reflection has been predicted in the normal metal/spin superconductor (NS) junctions based on the ferromagnetic graphene \cite{PhysRevB.95.104516}, where an incident spin-up electron is reflected as a spin-down electron and the spin $2\times(\hbar/2)$ is injected into the spin superconductor. This spin-flipped Andreev-like reflection leads to the spin current when a spin voltage is applied across the junction.

\begin{figure}[!tp]
\centerline{\includegraphics[width=1\linewidth]{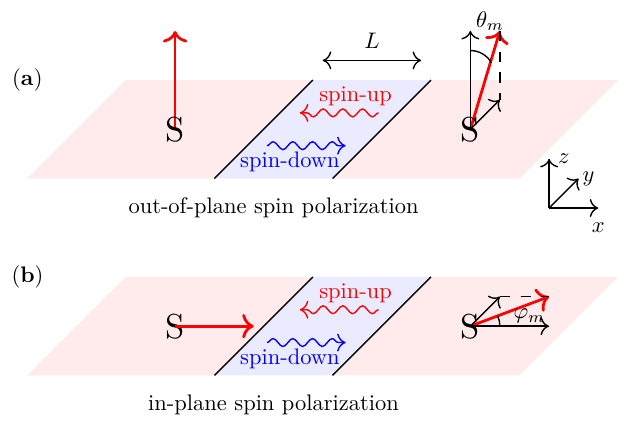}}
\caption{\label{fig:0}
Schematic of the spin Josephson junctions with the length of the normal region being $L$. The red arrows denote the directions of the spin polarization of the triplet exciton condensates. The red and blue wavy lines denote the left-propagating spin-up electron and the right-propagating spin-down electron, respectively.
(a) The spin Josephson junction with out-of-plane spin polarization. The spin polarization of the triplet exciton condensate in the left spin superconductor is along the $z$ axis and the spin polarization of the triplet exciton condensate in the right spin superconductor lies in the $y-z$ plane with the polar angle $\theta_m$. (b) The spin Josephson junction with in-plane spin polarization. The spin polarization of the triplet exciton condensate in the left spin superconductor is along the $x$ axis and the spin polarization of the triplet exciton condensate in the right spin superconductor lies in the $x-y$ plane with the azimuthal angle $\varphi_m$. }
\end{figure}

Inspired by this, in this paper we investigate the spin Josephson effect in the device consisting of two weakly coupled spin superconductors with noncollinear spin polarizations. Both the out-of-plane spin polarization configuration and the in-plane spin polarization configuration are considered, as shown in Figs. \ref{fig:0}(a) and \ref{fig:0}(b), respectively. We show that the repeated spin-flipped Andreev-like reflections at the NS interfaces lead to the Andreev-like bound states, which are responsible for the dissipationless spin supercurrent. This spin supercurrent is driven by the macroscopic phase difference $\phi_s$, which origins from a spin current flowing through the junction under the drive of an external device or from a variation of an external electric field thread the ring junction device \cite{PhysRevB.84.214501}. For the Josephson junctions with out-of-plane spin polarizations, the possible $\pi$-state spin supercurrent can be generated due to the Fermi momentum-splitting Andreev-like reflections at the NS interfaces. The anomalous spin supercurrent is absent and the phase of the spin supercurrent is determined by the Fermi level and the junction length. For the Josephson junctions with in-plane spin polarizations, the misorientation angle of the in-plane polarizations acts as an effective superconducting phase, leading to the anomalous spin supercurrent at the zero superconducting phase difference. Furthermore, the symmetry analysis shows that the appearance of the anomalous spin Josephson current is possible when the combined symmetry of the spin rotation and the time reversal is broken.

The remainder of the paper is organized as follows. The model Hamiltonian and the scattering approach are explained in detail in Sec.~\ref{sec:2}. The numerical results and discussions are presented in Sec.~\ref{sec:3}. Finally, we conclude in Sec.~\ref{sec:4}.

\section{Model and formulation}\label{sec:2}

\subsection{\textcolor{magenta}{Model Hamiltonian}}

The model Hamiltonian for the spin superconductor with the spin polarization along $z$ axis is given by \cite{PhysRevB.84.214501,PhysRevB.95.104516}
\begin{align}
\mathcal{H}_S=\begin{pmatrix}
\mathcal{H}_0+M & \Delta_s\tau_{z}\\ 
\Delta_s^*\tau_{z}& \mathcal{H}_0-M
\end{pmatrix},
\end{align}
where $M$ is the exchange split energy induced by the proximity to an insulating ferromagnetic layer \cite{PhysRevB.100.245148,PhysRevB.102.075443}, and $\Delta_s$ is the spin superconducting pair potential induced by the electron-hole Coulomb attraction interaction \cite{PhysRevB.84.214501}. The single-particle Hamiltonian $\mathcal{H}_0$ reads \cite{RevModPhys.80.1337,PhysRevLett.97.067007}
\begin{align}
\mathcal{H}_0=\hbar v_F(k_x\tau_x+\eta k_y\tau_y)-\mu,\label{eq:1}
\end{align}
where $\eta=+$ ($-$) for $\mathbf{K}$ ($\mathbf{K}'$) valley, $v_F$ is the Fermi velocity, $\mu$ is the Fermi energy, $k_x$ ($k_y$) is the wave vector in the $x$ ($y$) direction, and $\tau_{x,y,z}$ is the Pauli matrices in the pseudospin space. For the spin superconductor with the spin polarization 
\begin{align}
\mathbf{n}=(\sin\theta_m\cos\varphi_m,\sin\theta_m\sin\varphi_m,\cos\theta_m),
\end{align}
the Hamiltonian is given by
\begin{align}
\mathcal{H}_S(\mathbf{n})=\mathcal{R}_\mathbf{n}\mathcal{H}_S\mathcal{R}^\dagger_\mathbf{n},
\end{align}
where the spin rotation matrix is given by
\begin{gather}
\mathcal{R}_\mathbf{n}=\begin{pmatrix}
\cos\left(\frac{\theta_m}{2}\right)\tau_0 & e^{-i\varphi_m}\sin\left(\frac{\theta_m}{2}\right)\tau_0\\ 
e^{i\varphi_m}\sin\left(\frac{\theta_m}{2}\right)\tau_0 & \cos\left(\frac{\theta_m}{2}\right)\tau_0
\end{pmatrix},\label{eq:rr}
\end{gather}
with $\tau_0$ being the identity matrix in the pseudospin space, $\theta_m$ and $\varphi_m$ being the polar angle and the azimuthal angle, respectively.

We consider the spin superconductor/normal metal/spin superconductor junctions in the $x-y$ plane, where the spin polarizations of the two spin superconducting regions are noncollinear, as schematically shown in Fig.~\ref{fig:0}. The junction is described by the Bogoliubov-de Gennes (BdG) Hamiltonian \cite{de2018superconductivity}, which is given by 
\begin{gather}
\mathcal{H}_{BdG}=\mathcal{H}_L+\mathcal{H}_N+\mathcal{H}_R,\label{eq:hbdg}
\end{gather}
where $\mathcal{H}_L=\mathcal{H}_S(\mathbf{n}_L)\Theta(-x)$, $\mathcal{H}_R=\mathcal{H}_S(\mathbf{n}_R)\Theta(x-L)$ and $\mathcal{H}_N=\mathcal{H}_S|_{\Delta\rightarrow0}\Theta(x)\Theta(L-x)$ with $\mathbf{n}_{L,(R)}$ being the spin polarization for the left (right) spin superconductor and $\Theta(x)$ being the Heaviside step function. The superconducting pair potential is $\Delta_s=\Delta e^{-i\phi_s/2}$ for $x<0$ and $\Delta_s=\Delta e^{i\phi_s/2}$ for $x>L$, where $\phi_s$ is the phase difference across the junction.

\begin{figure*}[!tp]
\centerline{\includegraphics[width=\linewidth]{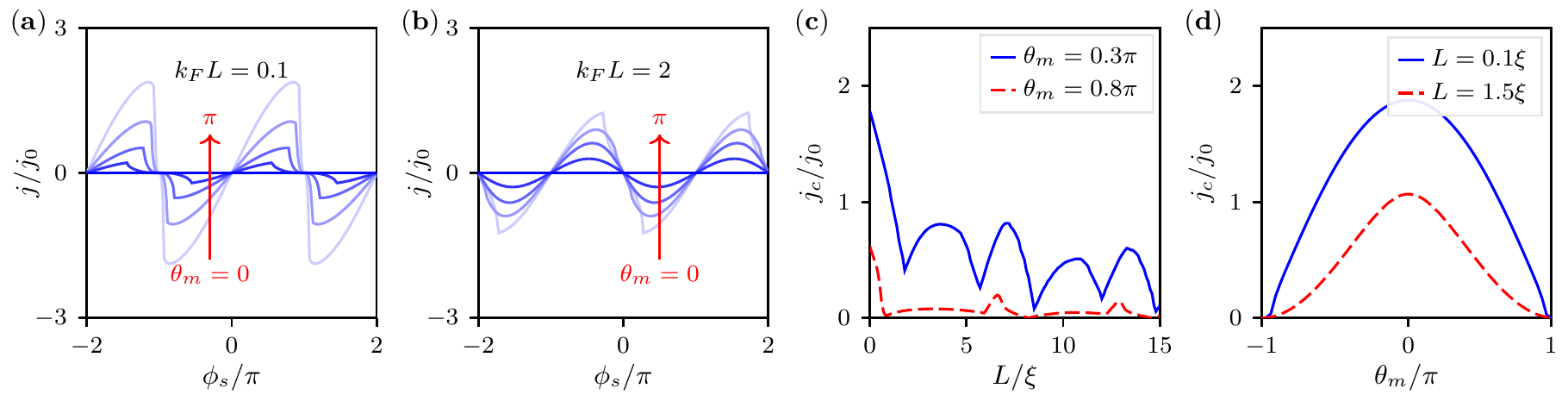}}
\caption{\label{fig:1}
(a, b) The current-phase relation for the Josephson junctions with the out-of-plane spin polarization. The red arrows indicate the curves with the increasing deviation of the relative angles $\theta_m$, which change from $0$ to $\pi$ with a step of $\pi/4$. (c) The critical spin Josephson current density $j_c$ as a function of $L$ for $\theta_m=0.3\pi$ (solid) and $\theta_m=0.8\pi$ (dashed). The Fermi level is $\mu=0.5\Delta$. (d) The critical spin Josephson current density $j_c$ as a function of $\theta_m$ for $L=0.1\xi$ (solid) and $L=1.5\xi$ (dashed). Both $j$ and $j_c$ are renormalized by $j_0=\mu\Delta W/2\pi\hbar v_F$.
}
\end{figure*}

\subsection{\textcolor{magenta}{Transfer matrix}}

The scattering states of the spin superconductor with spin polarization $\mathbf{n}$ can be obtained by 
\begin{align}
\psi^{(\mathbf{n})}_{\varsigma\varrho}
=\mathcal{R}_n
\begin{pmatrix}
e^{i\frac{\phi_s}{2}}e^{-i\frac{\varsigma\varrho\beta_\varrho}{2}}e^{i\frac{\varrho\gamma}{2}}\\ 
\varsigma\varrho e^{i\frac{\phi_s}{2}} e^{i\frac{\varsigma\varrho\beta_\varrho}{2}}e^{i\frac{\varrho\gamma}{2}}\\ 
e^{-i\frac{\phi_s}{2}}e^{-i\frac{\varsigma\varrho\beta_\varrho}{2}}e^{-i\frac{\varrho\gamma}{2}}\\ 
\varsigma\varrho e^{-i\frac{\phi_s}{2}} e^{i\frac{\varsigma\varrho\beta_\varrho}{2}}e^{-i\frac{\varrho\gamma}{2}}
\end{pmatrix}\times e^{i\varsigma\varrho k_\varrho x+ik_yy},\label{eq:states}
\end{align}
where $\varsigma=+$ $(-)$ for the right (left) propagating states, $\varrho=+$ $(-)$ for the electron-like (hole-like) quasiparticles, the longitudinal wave vector is $k_\varrho=\sqrt{(-M+i\varrho\sqrt{\Delta^2-E^2})^2/\hbar^2v_F^2-k_y^2}$ with $k_y$ and $E$ being the conserved transverse wave vector and the incident energy, respectively, $\beta_\varrho=\arctan(k_y/k_\varrho)$, $\gamma=\arccos(E/\Delta)$ for $E<\Delta$ and $\gamma=-i\mathrm{arccosh}(E/\Delta)$ for $E>\Delta$.

Due to the spin-flipped Andreev-like reflection at the NS interface \cite{PhysRevB.95.104516}, the incident and reflected states are connected by $\psi_{\bar\sigma}=\mathcal{S}\psi_\sigma$ with $\mathcal{S}$ being the transfer matrix, $\psi_\sigma$ ($\psi_{\bar{\sigma}}$) being the spin-$\sigma$ (spin-$\bar{\sigma}$) component in Eq.\ (\ref{eq:states}) ($\bar{\sigma}\equiv-\sigma$). The transfer matrices at the left and right interfaces can be directly obtained by matching the left and right propagating states in Eq.\ (\ref{eq:states}), respectively, which are given by
\begin{align}
\mathcal{S}_\nu=\frac{e^{i\varphi_\nu}\sin\left(\frac{\theta_\nu}{2}\right)+\cos\left(\frac{\theta_\nu}{2}\right)\mathcal{S}_\nu^{(0)}}{\cos\left(\frac{\theta_\nu}{2}\right)+e^{-i\varphi_\nu}\sin\left(\frac{\theta_\nu}{2}\right)\mathcal{S}_\nu^{(0)}},
\end{align}
with $\nu=+$ ($-$) for the left (right) interface. $\mathcal{S}_\nu^{(0)}$ is the Andreev-like transfer matrix for $\theta_m=0$, which is given by
\begin{gather}
\mathcal{S}_\nu^{(0)}=e^{-i\phi_\nu}\begin{pmatrix}
\cos\gamma &-i\nu\sin\gamma \\ 
i\nu\sin\gamma  & -\cos\gamma
\end{pmatrix}.
\end{gather}

In the normal region, the electron transfer matrix is readily obtained by \cite{PhysRevB.74.041401}
\begin{gather}
\mathcal{M}=\Lambda e^{ikL\tau_z}\Lambda,\\
\Lambda=\frac{1}{\sqrt{2\cos\alpha}}\begin{pmatrix}
e^{-i\alpha/2} &e^{i\alpha/2} \\ 
e^{i\alpha/2} & -e^{-i\alpha/2}
\end{pmatrix},
\end{gather}
where $k=k_F\cos\alpha$ is the longitudinal wave vector in the normal region with $k_F=\mu/\hbar v_F$ being the Fermi wave vector and $\alpha=\arcsin(\hbar v_Fk_y/\mu)\in[-\pi/2,\pi/2]$ is the incident angle.

The total spin Josephson current density can be obtained by \cite{PhysRevB.82.134516,brouwer1997anomalous}
\begin{align}
j=-k_BT\sum_{n,k_y}\frac{d}{d\phi_s}\ln \det\left(\mathmybb{1}-\mathcal{M}^{-1}\mathcal{S}_+^{-1}\mathcal{M}\mathcal{S}_-\right),\label{eq:jjT}
\end{align}
where $k_B$ is the Boltzmann constant, $T$ is the temperature, and the energy variable in the determinant is replaced by $i\omega_n$ with $\omega_n=(2n+1)\pi k_BT$ ($n\in\mathbb{Z}$).

\section{Results}\label{sec:3}
In the following calculations, we set $\Delta=\SI{1}{\meV}$ as the energy unit. The superconducting coherence length is $\xi=\hbar v_F/\Delta\approx\SI{360}{\nm}$, which is set as the length unit.

\begin{figure*}[!tp]
\centerline{\includegraphics[width=\linewidth]{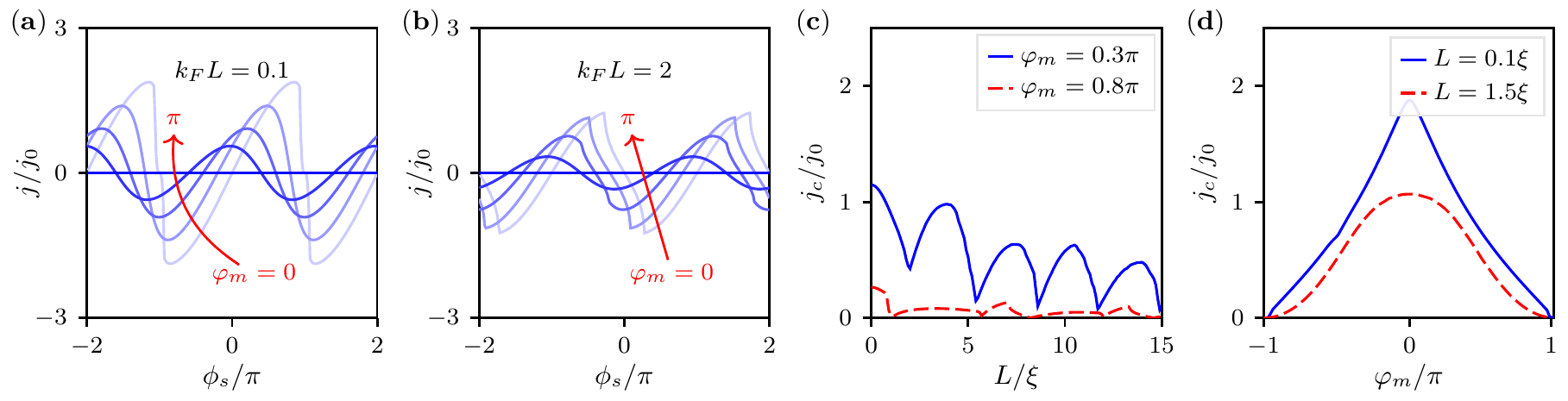}}
\caption{\label{fig:2}
(a, b) The current-phase relation for the Josephson junctions with the in-plane spin polarization. The red arrows indicate the curves with the increasing deviation of the relative angles $\varphi_m$, which change from $0$ to $\pi$ with a step of $\pi/4$. (c) The critical spin Josephson current density $j_c$ as a function of $L$ for $\varphi_m=0.3\pi$ (solid) and $\varphi_m=0.8\pi$ (dashed). The Fermi level is $\mu=0.5\Delta$. (d) The critical spin Josephson current density $j_c$ as a function of $\theta_m$ for $L=0.1\xi$ (solid) and $L=1.5\xi$ (dashed). Both $j$ and $j_c$ are renormalized by $j_0=\mu\Delta W/2\pi\hbar v_F$.}
\end{figure*}

\subsection{\textcolor{magenta}{Josephson junction with out-of-plane spin polarization}}

The Josephson junction with the out-of-plane spin polarization is schematically shown in Fig.\ \ref{fig:0}(a), where the spin polarization in the left superconducting region is along the $z$ axis and the spin polarization in the right superconducting region lies in the $y-z$ plane with the polar angle $\theta_m$. The Andreev-like transfer matrices at the NS interfaces are given by 
\begin{align}
\nonumber\mathcal{S}_\nu=&\frac{1}{\cos\theta_\nu\cos\phi_\nu+i\sin\phi_\nu}\times\\&\begin{pmatrix}
\sin\theta_\nu\cos\phi_\nu+\cos\gamma& -i\sin\gamma\\ 
i\sin\gamma& \sin\theta_\nu\cos\phi_\nu-\cos\gamma
\end{pmatrix},
\end{align}
where $\nu=+$ ($-$) for the right (left) interface, $\theta_+=\theta_m$, $\theta_-=0$, and $\phi_{\pm}=\pm\phi_s/2$. The Andreev-like bound states can be found from the same determinant in Eq.\ (\ref{eq:jjT}), which are given by
\begin{align}
\det\left(\mathmybb{1}-\mathcal{M}^{-1}\mathcal{S}_+^{-1}\mathcal{M}\mathcal{S}_-\right)=0,
\end{align}
leading to the Andreev-like levels
\begin{gather}
\left|\frac{E_{\pm}}{\Delta}\right|=\sqrt{\frac{Y^2-2XZ\pm\sqrt{Y^2-4Y^2Z\left(X+Z\right)}}{2\left(X^2+Y^2\right)}},\label{eq:ABS}\\
\nonumber X=2\sin^2\alpha-\left(3+\cos2\alpha\right)\cos(2kL),\\
\nonumber Y=-4\cos\alpha\sin(2kL),\\
\nonumber Z=\cos^2\alpha\big(\cos\phi_s+\cos\theta_m\cos\phi_s\\
\nonumber +\cos\theta_m+2\cos(2kL)-1\big).
\end{gather}
With the help of Eq.\ (\ref{eq:jjT}), the spin Josephson current density at $T=\SI{0}{\K}$ can be obtained by the $\phi_s$-dependent Andreev-like levels
\begin{align}
j(\phi_s)=-\sum_{\ell=\pm}\int\frac{\mu W}{2\pi\hbar v_F} \frac{dE_\ell}{d\phi_s}\cos\alpha d\alpha,\label{eq:jT0}
\end{align}
where $E_\ell$ is the positive root of Eq.\ (\ref{eq:ABS}) and $W$ is the width of the junction.

The supercurrent-phase relations at different $\theta_m$ are shown in Fig.\ \ref{fig:1}(a) for $k_FL=0.1$. The spin Josephson current decreases with the increasing of $\theta_m$, and drops to zero when the spin polarizations in the left and right superconducting regions are antiparallel, \textit{i.e.}, $\theta_m=\pi$. In this antiparallel configuration, the spin polarizations of the spin-triplet electron-hole condensations in the left and right spin superconductors are opposite, the Andreev-like process between two NS interfaces can not form a closed loop. Consequently, the $\phi_s$-dependent Andreev-like levels are abesnt, leading to the zero supercurrent.

For $k_FL=2$, the supercurrent is reversed and the $\pi$-state Josephson junction appears, as shown in Fig.\ \ref{fig:1}(b). This $0-\pi$ phase transition is attributed to the Fermi-momentum splitting Andreev-like reflections at the NS interfaces. At the Fermi surface, the right-propagating spin-up electron with the wave vector $k_F$ is converted into a left-propagating spin-down electron with the wave vector $-k_F$ due to the Andreev-like reflection, leading to the wave vector difference $\Delta k=2k_F$. An additional phase shift $2k_FL$ is accumulated in a closed Andreev loop. The total scattering phase acquired by the electrons in this closed Andreev loop is $2k_FL\pm\phi_s-2\arccos(E/\Delta)$~\cite{SatoshiKashiwaya_2000}, which should be quantized according to the quasiclassical Bohr-Sommerfeld quantization condition \cite{PhysRevD.25.1547,PhysRevLett.91.230402}, \textit{i.e.}, $2k_FL\pm\phi_s-2\arccos(E/\Delta)=2\pi n$ ($n\in\mathbb{Z}$). Consequently, the state of the junction can be controlled by the phase factor $\cos(k_FL)$. The spin supercurrent can be expressed as $j\propto\cos(k_FL)\sin\phi_s$. For $2n\pi+\pi/2<k_FL<2n\pi+3\pi/2$, the spin supercurrent is reversed.

The critical current defined by $j_c=\max|j(\phi_s)|$ as a function of the junction length $L$ is shown in Fig.\ \ref{fig:1}(c). Due to the additional phase $k_FL$, the critical current is proportional to $\cos(k_FL)$, which exhibits an oscillation behavior as the the junction length increases. For $\mu=0.5\Delta$ in Fig.\ \ref{fig:1}(c), the oscillation period can be estimated as $T_L=\pi/k_F=2\pi\xi$. The critical current as a function of the polar angle $\theta_m$ is shown in Fig.\ \ref{fig:1}(d). The $\theta_m$-dependent critical current is symmetric due to the rotation symmetry, \textit{i.e.}, $j_c(\theta_m)=j_c(-\theta_m)$. The critical current reaches its maximum value at $\theta_m=0$ and becomes zero at $\theta_m=\pm\pi$.

\subsection{\textcolor{magenta}{Josephson junction with in-plane spin polarization}}

The Josephson junction with in-plane spin polarization is schematically shown in Fig.\ \ref{fig:0}(b), where the spin polarization in the left superconducting region is along the $x$ axis and the spin polarization in the right superconducting region lies in the $x-y$ plane with the azimuthal angle $\varphi_m$. The Andreev-like transfer matrices are given by 
\begin{align}
\nonumber\mathcal{S}_\nu=&\frac{-ie^{i\varphi_\nu}}{\sin(\varphi_\nu+\phi_\nu)}\times\\&\begin{pmatrix}
\cos\left(\varphi_\nu+\phi_\nu\right)+\cos\gamma& -i\sin\gamma\\ 
i\sin\gamma& \cos\left(\varphi_\nu+\phi_\nu\right)-\cos\gamma
\end{pmatrix},
\end{align}
where $\nu=+$ ($-$) for the right (left) interface, $\varphi_+=\varphi_m$, $\varphi_-=0$, and $\phi_{\pm}=\pm\phi_s/2$. The Andreev-like levels are given by 
\begin{gather}
\left|\frac{E_{\pm}}{\Delta}\right|=\sqrt{\frac{\bar{Y}^2-2\bar{X}\bar{Z}\pm\sqrt{\bar{Y}^2-4\bar{Y}^2\bar{Z}\left(\bar{X}+\bar{Z}\right)}}{2\left(\bar{X}^2+\bar{Y}^2\right)}},\label{eq:ABS2}\\
\nonumber \bar{X}=8(3\cos(2kL)+2\cos2\alpha\cos^2(kL)-1),\\
\nonumber \bar{Y}=32\cos\alpha\sin(2kL),\\
\nonumber \bar{Z}=-4\cos^2\alpha\big(2\cos\varphi_m-\cos2\varphi_m+\cos\phi_s\\
\nonumber +2\cos(\varphi_m+\phi_s)+\cos(2\varphi_m+\phi_s)+4\cos(2kL)-1\big).
\end{gather}

The zero-temperature Josephson current obtained by Eq.\ (\ref{eq:jT0}) is shown in Fig.\ \ref{fig:2}(a) at $k_FL=0.1$. Due to the spin rotation symmetry, the supercurrent for the Josephson junctions with parallel in-plane spin polarization ($\varphi_m=0$) is the same as that for the Josephson junctions with parallel out-of-plane spin polarization ($\theta_m=0$). For the antiparallel in-plane polarization ($\varphi_m=\pi$), the Josephson current is absent due to the absence of the $\phi_s$-dependent Andreev-like levels. 

For $\varphi_m\neq0$ and $\varphi_m\neq\pi$, the anomalous Josephson current $j|_{\phi_s=0}$ appears, as shown in Fig.\ \ref{fig:2}(a). For simplicity, we present the spin Josephson current for the one-dimensional (1D) case with $k_y=0$:
\begin{align}
j_{1D}\propto\cos\left(k_FL\right)\cos^2\left(\frac{\varphi_m}{2}\right)\sin\left(\phi_s+\varphi_m\right),\label{eq:1d}
\end{align}
which can imply the qualitative properties of the two-dimensional Josephson junctions. Eq.\ (\ref{eq:1d}) shows that the anomalous Josephson current at $\phi_s=0$ is proportional to $\cos(k_FL)\cos^2(\varphi_m/2)\sin\varphi_m$. The Josephson current can be reversed by the additional phase factor $\cos(k_FL)$. For $k_FL=2$, the reversed Josephson current obtained from Eq.\ (\ref{eq:jT0}) is shown in Fig.\ \ref{fig:2}(b).

The critical currents $j_c$ as a function of the junction length $L$ and the azimuthal angle $\varphi_m$ are shown in Figs.\ \ref{fig:2}(c) and \ref{fig:2}(d), respectively. The $L$-dependent oscillation is determined by the additional phase factor $\cos(k_FL)$ due to the Fermi-moment splitting. For a fixed $L$, the $\varphi_m$-resolved $j_c$ is symmetric, as shown in Fig.\ \ref{fig:2}(d).

\begin{figure}[!bp]
\centerline{\includegraphics[width=\linewidth]{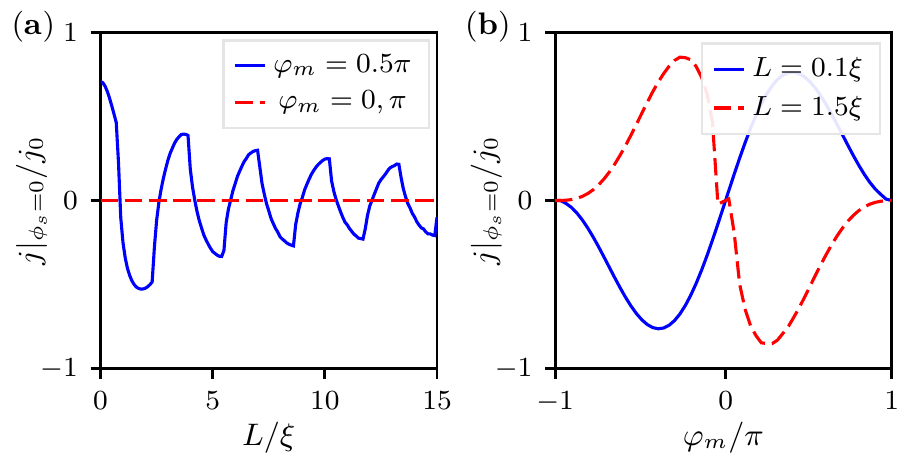}}
\caption{\label{fig:3}
(a) The anomalous spin Josephson current density $j|_{\phi_s=0}$ as a function of $L$ for $\varphi_m=0.5\pi$ (solid) and $\varphi_m=0,\pi$ (dashed). The Fermi level is $\mu=\Delta$. (b) The anomalous spin Josephson current $j|_{\phi_s=0}$ as a function of $\varphi_m$ for $L_m=0.1\xi$ (solid) and $L=1.5\xi$ (dashed). $j|_{\phi_s=0}$ is renormalized by $j_0=\mu\Delta W/2\pi\hbar v_F$.
}
\end{figure}

The anomalous Josephson current $j|_{\phi_s=0}$ as a function of $L$ is shown in Fig.\ \ref{fig:3}(a). For $\varphi_m=0$ and $\varphi_m=\pi$, the anomalous Josephson effect is absent. For $\varphi_m\neq0$ and $\varphi_m\neq\pi$, $j|_{\phi_s=0}$ oscillates with the increasing of $L$ and has the same oscillation period as $j_c$. For $\mu=\Delta$ in Fig.\ \ref{fig:3}(a), the oscillation period is about $\pi\xi$. $j|_{\phi_s=0}$ as a function of $\varphi_m$ is shown in Fig.\ \ref{fig:3}(b). The $\varphi_m$-resolved anomalous Josephson current is asymmetric and satisfies the relation
\begin{align}
j|_{\phi_s=0}(\varphi_m)=-j|_{\phi_s=0}(-\varphi_m),
\end{align}
due to the symmetry of the BdG Hamiltonian, which is explained in detail in Sec.\ \ref{Symmetry}.

\subsection{\textcolor{magenta}{Symmetry analysis}}\label{Symmetry}
The Hamiltonian of the spin superconductor with out-of-plane spin polarization can be written as
\begin{align}
\mathcal{H}^\eta_S=h_\eta(k)+\mathcal{H}'(\phi_s,\theta_m),
\end{align}
where $\eta=\pm$ is the valley index, the $k$-dependent term $h_\eta(k)=\hbar v_F(k_x\tau_x+\eta k_y\tau_y)$ is the Hamiltonian of the pristine graphene for the $\eta$ valley and the $k$-independent term is given by
\begin{align}
\nonumber\mathcal{H}'(\phi_s,\theta_m)=&M\cos\theta_m\sigma_z\tau_0+M\sin\theta_m\sigma_x\tau_0\\\nonumber&+\Delta\cos\theta_m\cos\phi_s\sigma_x\tau_z+\Delta\sin\theta_m\cos\phi_s\sigma_z\tau_z\\&-\Delta\sin\phi_s\sigma_y\tau_z.
\end{align}
Under the time-reversal transformation, the Hamiltonian of the pristine graphene remains unchanged
\begin{align}
\nonumber h_{\eta}(k)&\rightarrow\mathcal{T}h_{\bar{\eta}}(-k)\mathcal{T}^{-1}\\&\nonumber=\hbar v_F(k_x\tau_x+\eta k_y\tau_y)\\
&=h_{\eta}(k),
\end{align}
where $\bar{\eta}=-\eta$, $\mathcal{T}=i\sigma_y\tau_z\mathcal{C}$ is the time-reversal operator with $\mathcal{C}$ being the complex conjugation. For the $k$-independent Hamiltonian $\mathcal{H}'(\phi_s,\theta_m)$, considering the combined symmetry $\mathcal{P}=\sigma_y\mathcal{T}$ with $\sigma_y$ being the spin rotation symmetry and $\mathcal{T}$ being the time-reversal symmetry, one finds that 
\begin{align}
\mathcal{P}\mathcal{H}'(\phi_s,\theta_m)\mathcal{P}^{-1}=\mathcal{H}'(-\phi_s,\theta_m).
\end{align}
Due to the spin degeneracy in pristine graphene, $h_\eta(k)$ remains unchanged under $\mathcal{P}$. Consequently, with the help of Eq.\ (\ref{eq:hbdg}), the BdG Hamiltonian for the Josephson junctions with the out-of-plane spin polarization holds the relation
\begin{align}
\mathcal{P}\mathcal{H}_{BdG}(\phi_s,\theta_m)\mathcal{P}^{-1}=\mathcal{H}_{BdG}(-\phi_s,\theta_m). \label{eq:symmetry1}
\end{align}
The spin Josephson current density can be obtained via the thermodynamic relation \cite{PhysRev.187.556,PhysRevLett.67.3836}
\begin{align}
j(\phi_s)=-k_BT\partial_{\phi_s}\ln\Tr[e^{-\mathcal{H}_{BdG}(\phi_s)/k_BT}].\label{eq:freeenergy}
\end{align}
By substituting Eq.\ (\ref{eq:symmetry1}) into Eq.\ (\ref{eq:freeenergy}), one finds that the spin Josephson current holds the relation  
\begin{align}
j(\phi_s,\theta_m)=-j(-\phi_s,\theta_m).
\end{align}
leading to the absence of the anomalous spin Josephson current, \textit{i.e.}, $j|_{\phi_s=0}=0$.

For the spin superconductor with in-plane spin polarization, the Hamiltonian can be written as
\begin{align}
\mathcal{H}^\eta_S=h_\eta(k)+\mathcal{H}''(\phi_s,\varphi_m),
\end{align}
with 
\begin{align}
\nonumber\mathcal{H}''(\phi_s,\varphi_m)=&M\cos\varphi_m\sigma_x\tau_0+M\sin\varphi_m\sigma_y\tau_0\\&\nonumber-\Delta\sin(\varphi_m+\phi_s)\cos\varphi_m\sigma_y\tau_z\\&\nonumber+\Delta\sin(\varphi_m+\phi_s)\sin\varphi_m\sigma_x\tau_z\\&-\Delta\cos(\varphi_m+\phi_s)\sigma_z\tau_z.
\end{align}
The symmetry $\mathcal{P}$ is broken and the BdG Hamiltonian holds the relation
\begin{align}
\mathcal{P}\mathcal{H}_{BdG}(\phi_s,\varphi_m)\mathcal{P}^{-1}=\mathcal{H}_{BdG}(-\phi_s,-\varphi_m).
\end{align}
With the help of Eq.\ (\ref{eq:freeenergy}), one finds that the spin Josephson current holds the relation 
\begin{align}
j(\phi_s,\varphi_m)=-j(-\phi_s,-\varphi_m),
\end{align}
leading to the presence of the anomalous spin Josephson current. The anomalous Josephson current is an odd function of $\varphi_m$ satisfying $j|_{\phi_s=0}(\varphi_m)=-j|_{\phi_s=0}(-\varphi_m)$, as shown in Fig.\ (\ref{fig:3})(b).

In fact, the anomalous Josephson effect is possible when the symmetries $\mathcal{T}$, $\mathcal{D}$ and $\mathcal{P}$ are all broken \cite{PhysRevB.82.125305,PhysRevB.93.155406}, where $\mathcal{D}=\sigma_yR_{xz}\mathcal{T}$ is another combined symmetry with $R_{xz}$ being the mirror reflection about the $x-z$ plane. We note that the symmetries $\mathcal{T}$ and $\mathcal{D}$ are always broken in our model for both the out-of-plane spin polarization and the in-plane spin polarization. The combined symmetry $\mathcal{P}$ plays a key role in our model: For the Josephson junction with out-of-plane spin polarization, the absence of the anomalous Josephson current ($j|_{\phi_s=0}=0$) is protected by $\mathcal{P}$. For the Josephson junction with in-plane spin polarization, the breaking of $\mathcal{P}$ leads to the presence of the anomalous spin Josephson current.\\ \\

\section{Conclusions}\label{sec:4}
To conclude, we theoretically investigate the spin Josephson effect in the spin superconductor/normal metal/spin superconductor junctions, where the spin polarizations of the left and right spin superconducting regions are noncollinear. For the Josephson junctions with out-of-plane spin polarization, the possible $\pi$-state spin supercurrent can be generated due to the Fermi momentum-splitting Andreev-like reflection at the normal metal/spin superconductor interface. For the Josephson junctions with in-plane spin polarization, the anomalous spin supercurrent appears and is driven by the misorientation angle of the in-plane spin polarization. The symmetry analysis shows that the appearance of the anomalous spin Josephson current is possible when the combined symmetry of the spin rotation and the time reversal is broken.

\section*{Acknowledgements}
This work is supported by the National Key R\&D Program of China (Grant No. 2022YFA1403601).

\bibliographystyle{elsarticle-num} 
% \bibliography{mybib}

\end{document}